\documentclass[journal]{IEEEtran}
\usepackage{amssymb}
\usepackage{amsmath}

\usepackage{graphicx}
\usepackage{cite}
\usepackage{txfonts}
\usepackage{mathrsfs}
\usepackage{fontenc}
\usepackage[binary-units]{siunitx}
\DeclareSIUnit{\bps}{bps}
\usepackage[caption=false,font=footnotesize]{subfig}
\usepackage{tabularx}
\usepackage{multirow}
\usepackage{diagbox}
\usepackage{algorithm}
\usepackage{algorithmic}
\usepackage{bbm}
\usepackage{xcolor}

\begin{document}
\title{Performance Impact of Channel Aging and Phase Noise on Intelligent Reflecting Surface}

\author{Wei~Jiang,~\IEEEmembership{Senior~Member,~IEEE,}
        and Hans~Dieter~Schotten
\thanks{\textit{Corresponding author: Wei Jiang (e-mail: wei.jiang@dfki.de)}}
}

\markboth{IEEE XXX Letters,~Vol.~x, No.~x, YY~202X}%
{Jiang \MakeLowercase{\textit{et al.}}: Performance Impact of Channel Aging and Phase Noise}
\maketitle

\begin{abstract}
This letter aims to clarify the impact of channel aging and phase noise on the performance of intelligent reflecting surface-aided wireless systems. We first model mathematically the outdated channel state information (CSI) due to Doppler shifts and phase noise stemming from hardware impairment. Then, a closed-form expression of achievable spectral efficiency under noisy and aged CSI is theoretically derived. Some typical simulation results to numerically demonstrate the performance impact are illustrated.
\end{abstract}

\IEEEpeerreviewmaketitle

\section{Introduction}

\IEEEPARstart{R}{ecently}, intelligent reflecting surface (IRS) attracted much interest from academia and industry. Through smartly shaping propagation environment \cite{Ref_wu2019intelligent}, it enables a new degree of freedom for the forthcoming sixth-generation (6G) system.
The achievement of IRS potential heavily relies on judiciously adjusting its reflecting phases in terms of instantaneous channel state information (CSI). Although some prior works studied the effect of estimation errors  \cite{Ref_you2020channel} and the feasibility of using statistical CSI \cite{Ref_han2019large} to avoid the reliance on instantaneous CSI,  most optimization works assume that the CSI is perfectly known and keeps constant from the acquisition time to its actual usage \cite{Ref_wu2019intelligent}. From a practical point of view, however, the estimated CSI may substantially differ from the actual CSI when using the reconfigured phases for signal reflection. Utilizing outdated phase shifts on the surface may severely deteriorate the system performance and even overwhelm the achievable gain of applying IRS.

Many researchers mentioned that channel aging may impose a significant performance loss on IRS \cite{Ref_renzo2020smart}. To the best knowledge of the authors, however, none of them provided either a theoretical analysis or a numerical result. There are indeed many previous works about the effect of channel aging. But the conventional wireless techniques such as \cite{Ref_jiang2021impactcellfree} merely adjust transmission parameters based on the \textit{amplitude} or power gain of a fading channel, as discussed in \cite{Ref_myOJCOMS}, whereas the \textit{phase} of channel is unused and simply ignored. Consequently, phase-based aging analysis is missing in the literature by far since there is no driver before. As a phase-based adaptive technique, IRS imposes an inimitable demand to analyze aging from the perspective of phase. That is still an open issue.
In addition to channel imperfection, phase noise due to hardware imperfection also causes the mismatch of reflecting phases. Some works \cite{Ref_badiu2020communication} discussed the effect of phase noise due to the lack of high-precision configuration of IRS reflectors (we name it \textit{reflector's phase noise} hereinafter). However, the phase noise stemmed from the oscillator of base stations (BS) or user equipment (UE) (we call it \textit{oscillator's phase noise}) is still an open issue in the field of IRS.

To fill this gap, this letter aims to clarify the impact of channel aging and phase noise on IRS-aided systems. The structure is summarized as follows: Section II presents an IRS system model under channel aging and phase noise. Section III analyzes channel and hardware impairments, and Section IV provides a unified model of aged and noisy CSI. Next, a closed-form expression of achievable spectral efficiency in the presence of aged and noisy CSI  is derived in Section V. In Section VI, performance evaluation is illustrated. The conclusions are drawn in Section VII.

\section{System Model}

\begin{figure}[!t]
    \centering
    \includegraphics[width=0.39\textwidth]{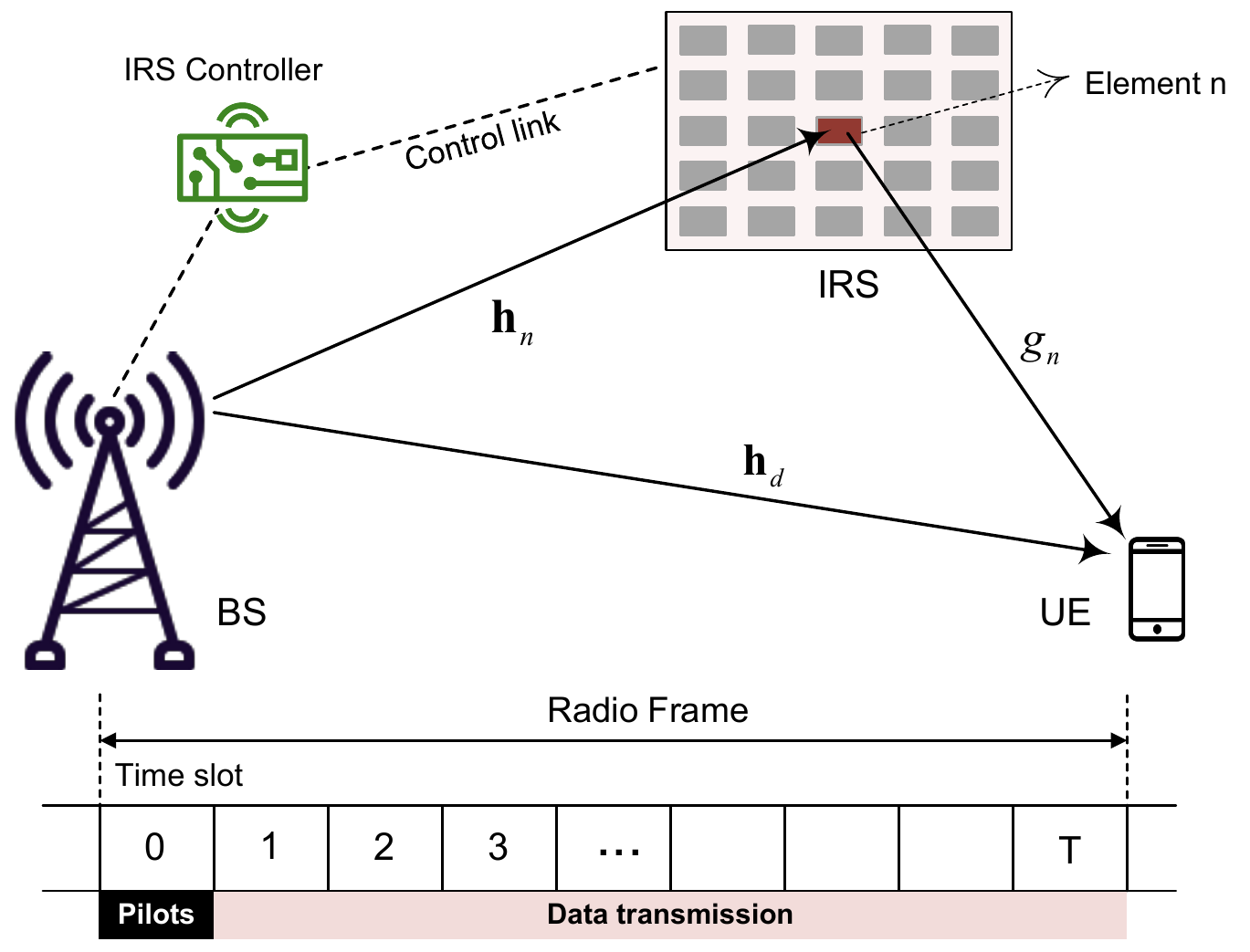}
    \caption{Schematic diagram of an IRS-aided system and the frame structure.  }
    \label{fig:system}
\end{figure}

Consider a single-cell multi-input single-output communications system comprising an $N_b$-antenna BS, a single-antenna UE, and an IRS with $N$ reflecting elements \cite{Ref_wu2019intelligent}. The IRS is equipped with a smart controller to dynamically adjust the phase shift of each reflector in terms of the instantaneous CSI acquired through periodic estimation.
Due to high path loss, the signals reflected by the IRS twice or more are negligible.  As illustrated in \figurename \ref{fig:system}, a radio frame is comprised of $\mathcal{T}+1$ time slots, and the duration of each slot is $T_s$. Since the IRS is passive, time-division duplexing operation with channel reciprocity is usually adopted to simplify channel estimation. The first time slot is dedicated to uplink training, while the subsequent $\mathcal{T}$ slots are used for data transmission. Both passive beamforming at the IRS and active beamforming at the BS are determined by the estimated CSI at slot $0$, and keep unchanged for the remaining $\mathcal{T}$ slots.
Because the impact of channel aging and phase noise is equivalent regardless of downlink or uplink, this letter only focuses on the downlink, whereas skipping the uplink. But the theoretical analysis can be applied straightforwardly.

Without losing generality, we assume that the channel response and phase noise vary in slot-wise but keep constant within a slot.
Note that a pilot signal experiences the oscillator phase noise at the BS and UE, like a kind of channel response. To characterize the theoretical analysis, we assume that the estimated CSI is perfect by neglecting estimation errors.  Therefore, the estimated CSI consists of three components: the phase noise induced at the transmitter, channel gain, and phase noise at the receiver. We can write
\begin{equation} \label{EQN_CSIhd0}
    \mathfrak{h}_{dn_b,0}=e^{j\phi_{0}}h_{dn_b,0} e^{j\psi_{0}}=h_{dn_b,0} e^{j(\phi_{0}+\psi_{0})}
\end{equation}
to denote the estimated CSI of the channel from  BS antenna $n_b=1,\ldots,N_b$ to the UE, where $h_{dn_b,0}$ represents instantaneous channel gain during the uplink training, $\phi_{0}$ and $\psi_{0}$ denote the phase noise of the BS and UE oscillators at slot $0$, respectively. As time elapses, the channel gain changes to $h_{dn_b,t}$ at slot $t$, and the oscillator phase noise of the BS and UE become $\phi_{t}$ and $\psi_{t}$, respectively. Thus, the actual CSI at slot $t$ is given by
\begin{equation} \label{EQN_CSIhdt}
\mathfrak{h}_{dn_b,t}=h_{dn_b,t} e^{j(\phi_{t}+\psi_{t})},\:\:\forall t=1,2,\ldots, \mathcal{T}.
\end{equation}
The estimated CSI $\mathfrak{h}_{dn_b,0}$ is an outdated version of the actual CSI $\mathfrak{h}_{dn_b,t}$. Their statistical relationship will be mathematically modelled in the subsequent section.

Since a reflecting surface is passive without any oscillator, only transmitter phase noise exists in the BS-IRS link.
As a result, the estimated CSI between the $n_b^{th}$ BS antenna and the $n^{th}$ IRS element equals
$\mathfrak{h}_{nn_b,0}= h_{nn_b,0} e^{j\phi_{0}}$. The actual CSI at time slot $t$ is given by
\begin{equation} \label{EQN_CSIhnt}
\mathfrak{h}_{nn_b,t} = h_{nn_b,t} e^{j\phi_{t}}, \:\:\forall t=1,2,\ldots,\mathcal{T}.
\end{equation}
Similarly, we merely need to consider receiver phase noise in the IRS-UE link. The estimated and actual CSI between the $n^{th}$ IRS element and UE are expressed as $\mathfrak{g}_{n,0}= g_{n,0} e^{j\psi_{0}}$  and
\begin{equation} \label{EQN_CSIgnt}
\mathfrak{g}_{n,t} = g_{n,t} e^{j\psi_{t}}, \:\:\forall t=1,2,\ldots,\mathcal{T},
\end{equation} respectively.

The BS applies linear beamforming with a transmit vector $\mathbf{w}\in \mathbb{C}^{N_b\times 1}$, satisfying $\|\mathbf{w}\|^2\leqslant 1$, where $\|\cdot\|$ represents the Euclidean norm of a complex vector. Then, the \textit{discrete-time} baseband equivalent signal received by the UE is
\begin{equation} \label{eqn_systemModel}
    r_t=\sqrt{P_d} \Biggl( \sum_{n=1}^N \mathfrak{g}_{n,t} c_n \mathbf{h}^T_{n,t} + \mathbf{h}_{d,t}^T \Biggr)\mathbf{w} \mathbf{s}_t + \mathbf{n}_t, \:\:\forall t=1,2,\ldots,\mathcal{T},
\end{equation}
where the superscript $(\cdot)^T$ means the transpose of a matrix or vector, $\mathbf{s}_t=\left[s_{1,t},\ldots,s_{T_s,t}\right]\in \mathbb{C}^{1\times T_s}$ denotes the vector of $T_s$ transmitted symbols at time slot $t$, with $\mathbb{E}\left[|s|^2\right]=1$, $P_d$ expresses the power constraint of the BS, and $\mathbf{n}_t$ is a vector of additive white Gaussian noise (AWGN) with zero mean and variance $\sigma_n^2$, i.e., $\mathbf{n}_t\sim \mathcal{CN}(\mathbf{0},\sigma_n^2\mathbf{I}_{T_s})$. Meanwhile, we write
\begin{equation} \label{EQNIRS_VectorBSUE}
    \mathbf{h}_{d,t}=[\mathfrak{h}_{d1,t},\mathfrak{h}_{d2,t},\ldots,\mathfrak{h}_{dN_b,t}]^T\in \mathbb{C}^{N_b\times 1}
\end{equation} to denote the overall BS-UE channel vector, and
\begin{equation} \label{EQNirs_vecIRSUE}
    \mathbf{h}_{n,t}=[\mathfrak{h}_{n1,t},\mathfrak{h}_{n2,t},\ldots,\mathfrak{h}_{nN_b,t}]^T\in \mathbb{C}^{N_b\times 1}
\end{equation}
to denote the channel vector from the BS to the $n^{th}$ reflecting element at slot $t$.

The reflection coefficient of the $n^{th}$ IRS element is expressed as $c_n=\alpha_n e^{j\theta_n}$, with a phase shift $\theta_n\in [0,2\pi)$ and attenuation $\alpha_n\in [0,1]$. As revealed by \cite{Ref_wu2019intelligent}, the optimal attenuation is $\alpha_n=1$, $\forall n=1,2,\ldots,N$ to maximize the received power and simplify hardware implementation.
Let
\begin{align} \label{EQNIRS:VectorIRSUE}
    \mathbf{g}_t&=\left[\mathfrak{g}_{1,t},\mathfrak{g}_{2,t},\ldots,\mathfrak{g}_{N,t}\right]^T\\
    \boldsymbol{\Theta}&=\mathrm{diag}\left\{e^{j\theta_1},e^{j\theta_2},\ldots,e^{j\theta_N}\right\},
\end{align}
 and  $\mathbf{H}_t\in \mathbb{C}^{N\times N_b}$ to denote the BS-IRS channel matrix, where the $n^{th}$ row of $\mathbf{H}_t$ equals to $\mathbf{h}_{n,t}^T$, \eqref{eqn_systemModel} can be rewritten in matrix form as
\begin{equation}
    r_t= \sqrt{P_d}\biggl(\mathbf{g}_t^T \boldsymbol{\Theta} \mathbf{H}_t +\mathbf{h}_{d,t}^T\biggr)\mathbf{w} \mathbf{s}_t + \mathbf{n}_t.
\end{equation}

\section{Channel and Hardware Impairments}
The performance gain of IRS-aided systems heavily relies on the accurate manipulation of reflection phases. However, channel and hardware impairments cause the mismatch of reflection phases. We will mathematically model the channel aging raised by \textit{Doppler shifts} and  the \textit{phase noise} due to imperfect transceiver oscillators or low-precision IRS reflectors, as the basis for analyzing the performance impact.

\subsection{Channel Aging} \label{SubSec:Aging}

The movement of users or their surrounding scatters leads to a time-varying channel. For the sake of simplicity, we can ignore different subscripts under independent and identically-distributed (\emph{i.i.d.}) channels. Hence, we generally denote the channel gain at the instant of uplink training by $h_0$, which is an outdated version of the actual value $h_t$ during data transmission at slot $t$. A metric known as correlation coefficient is used to quantify the channel aging \cite{Ref_jiang2016robust}, i.e., $\rho_t=\frac{\mathbb{E}\left[h_t h_0^*\right]}{\sqrt{\mathbb{E}[|h_t|^2] \mathbb{E}[|h_0|^2]}}$,
where $\mathbb{E}\left[ \cdot \right]$ stands for mathematical expectation, $(\cdot)^*$ means complex conjugate.  We have
\begin{equation} \label{eqn:outdatedCSI}
h_t=\left ( \rho_t h_0 + \varepsilon \sqrt{1-\rho_t^2}  \right )
\end{equation}
with an innovation component $\varepsilon$, which is a standard complex normal random variable, i.e., $\varepsilon \sim \mathcal{CN}(0,1)$.
Under the classical Doppler spectrum of the Jakes' model,  the correlation coefficient takes the value $\rho_t=J_0(2\pi f_d tT_s)$, where $f_d$ means the maximal Doppler shift, $tT_s$ stands for the delay between the outdated and actual CSI, and $J_0(\cdot)$ represents the $zeroth$ order Bessel  function  of the first kind. In particular, $f_d$ can be computed by $
    f_d=\frac{f_c v}{c}=\frac{v}{\lambda}$,
where $v$ denotes the velocity of a moving object, $c$ is the speed of light in free space, and $\lambda$ represents the wavelength of carrier frequency $f_c$.

\subsection{Oscillator Phase Noise} \label{SubSec:Oscillator}
Due to oscillator imperfection, a transmitted signal suffers from phase noise during the up-conversion from baseband to passband, and \textit{vice versa} at the receiver. Such phase noise is not only random but also time-varying.  Unlike a non-synchronous setup where each antenna has its own oscillator, e.g., a distributed antenna system, an IRS-aided system generally employs a BS with co-located antennas. Hence, our analysis only considers the synchronous operation where all co-located antennas at the BS share a common oscillator.

Utilizing a well-established  Wiener process \cite{Ref_Papazafeiropoulos2017impact},  the phase noise of the BS and UE at time slot $t$ can be modeled as
\begin{equation} \label{Eqn:IRS:phaseNoise} \begin{cases}
\phi_{t}= \phi_{0} + \sum_{\tau=1}^t \triangle \phi_{\tau}= \phi_{0} + \triangle \Phi_{t} \\
\psi_{t}=\psi_{0}+\sum_{\tau=1}^t \triangle \psi_{\tau}=\psi_{0}+ \triangle \Psi_{t},\end{cases}
\end{equation}
where $\phi_{0}$ and $\varphi_{0}$ represent the phase noise of the BS and UE at slot $0$, respectively. The incremental noise $\triangle \phi_{t}= \phi_{t}-\phi_{t-1}$ and $\triangle \psi_{t}= \psi_{t}-\psi_{t-1}$ are normal random variables, i.e., $\triangle \phi_{t}\sim \mathcal{N}(0,\sigma_{\phi_{}}^{2})$ and $\triangle \psi_{t} \sim \mathcal{N}(0,\sigma_{\psi_{}}^{2})$, where $\sigma_{i}^{2}=4\pi^{2}f_{\mathrm{c}} c_{i}T_{\mathrm{s}}$, $\forall i=\phi, \psi$ with  the oscillator-dependent constant  $c_{i}$.
For simple notation, we use $\triangle \Phi_{t}=\sum_{\tau=1}^t \triangle \phi_{\tau}$ and $\triangle \Psi_{t}=\sum_{\tau=1}^t \triangle \psi_{\tau}$ to denote the accumulated phase noise from time slot $1$ to $t$.  It is easy to derive that $\triangle \Phi_{t} \sim \mathcal{N}(0,t\sigma_{\phi_{}}^{2})$ and $\triangle \Psi_{t} \sim \mathcal{N}(0,t\sigma_{\varphi_{}}^{2})$.  

\subsection{Reflector Phase Noise}
In a practical system, high-precision configuration of the reflecting elements is unfeasible. Mathematically, the resulting phase shift of the $n^{th}$ IRS element is $\hat{\theta}_n=\theta_n+\tilde{\theta}_n$, where $\theta_n$ is the desired value and $\tilde{\theta}_n$ stands for phase noise. As \cite{Ref_wang2021outage}, $\tilde{\theta}_n$, $\forall n$ can be modeled by mutually independent and identical Von Mises random variables with zero mean and concentration parameter $\kappa$. Its probability density function is given by
\begin{equation}
    f_{\tilde{\theta}_n}(x) = \frac{e^{\kappa \cos{x}}}{2\pi I_0(\kappa)}, \:\:\:\: -\pi \leqslant x < \pi,
\end{equation}
where $I_0(\cdot)$ denotes the $zeroth$ order modified Bessel  function  of the first kind. For ease of analysis, we define a matrix of phase errors as $\tilde{\boldsymbol{\Theta}}=\mathrm{diag}\left\{e^{j\tilde{\theta}_1},e^{j\tilde{\theta}_2},\ldots,e^{j\tilde{\theta}_N}\right\}$.

\section{Model of Aged and Noisy CSI}
According to \eqref{eqn:outdatedCSI} and \eqref{Eqn:IRS:phaseNoise}, we can project the channel gain and phase noise at slot $t$ given the estimates at slot $0$.
Thus, substitute \eqref{eqn:outdatedCSI} and \eqref{Eqn:IRS:phaseNoise} into \eqref{EQN_CSIhdt} to obtain the aged and noisy CSI at slot $t$ as
\begin{align} \nonumber \label{EQNIRS:CSIatslott}
\mathfrak{h}_{dn_b,t}&=\underbrace{\left ( \rho_t h_{dn_b,0} + \varepsilon_{n_b} \sqrt{1-\rho_t^2}  \right )}_{\mathrm{Aged\:Channel\:Gain:\:\eqref{eqn:outdatedCSI}}} \underbrace{e^{j\left(\phi_{0}+\psi_{0}+\triangle \Phi_{t}+\triangle \Psi_{t}\right)}}_{\mathrm{Varying\:Phase\:Noise:\:\eqref{Eqn:IRS:phaseNoise}}}\\
&= \rho_t \mathfrak{h}_{dn_b,0} e^{j\left(\triangle \Phi_{t}+\triangle \Psi_{t}\right)}+ \varepsilon_{n_b} \sqrt{1-\rho_t^2}   e^{j\left(\phi_{0}+\psi_{0}+\triangle \Phi_{t}+\triangle \Psi_{t}\right)}.
\end{align}
Replacing the entries in \eqref{EQNIRS_VectorBSUE} with \eqref{EQNIRS:CSIatslott}, we get the relationship between the actual and outdated channel vectors as
\begin{equation} \label{EQNIRS:Vecdirectlink}
    \mathbf{h}_{d,t}=\rho_t\mathbf{h}_{d,0}e^{j\left(\triangle \Phi_{t}+\triangle \Psi_{t}\right)}+ \boldsymbol{\varepsilon}_d \sqrt{1-\rho_t^2}   e^{j\left(\phi_{0}+\psi_{0}+\triangle \Phi_{t}+\triangle \Psi_{t}\right)}
\end{equation}
with the definition of a vector of innovation components as $\boldsymbol{\varepsilon}_d=[\varepsilon_1,\varepsilon_2,\ldots,\varepsilon_{N_b}]^T$.

Applying \eqref{eqn:outdatedCSI} and \eqref{Eqn:IRS:phaseNoise}, \eqref{EQN_CSIgnt} can be rewritten to obtain the actual CSI between the $n^{th}$ IRS element and the UE in the presence of channel aging and phase noise as
\begin{align} \nonumber \label{EQNIRS:AgedNoisyCSI}
\mathfrak{g}_{n,t} &= \left ( \rho_t g_{n,0} + \varepsilon_n \sqrt{1-\rho_t^2}  \right ) e^{j(\psi_{0}+\triangle \Psi_{t})}\\
&=  \rho_t \mathfrak{g}_{n,0} e^{j\triangle \Psi_{t}}+ \varepsilon_n \sqrt{1-\rho_t^2}   e^{j(\psi_{0}+\triangle \Psi_{t})}.
\end{align}
 Analogue to \eqref{EQNIRS:Vecdirectlink}, we can substitute \eqref{EQNIRS:AgedNoisyCSI} into \eqref{EQNIRS:VectorIRSUE} to obtain
\begin{equation}
    \mathbf{g}_{t}=\rho_t\mathbf{g}_{0}e^{j\triangle \Psi_{t}}+ \boldsymbol{\varepsilon}_g \sqrt{1-\rho_t^2}   e^{j\left(\psi_{0}+\triangle \Psi_{t}\right)},
\end{equation}
with $\boldsymbol{\varepsilon}_g=[\varepsilon_1,\ldots,\varepsilon_{N}]^T$.
The BS-IRS link is generally a line of sight (LOS) due to favourable locations of the BS and IRS. Its channel response can be regarded as time-invariant and therefore does not suffer from channel aging, namely $h_{nn_b,t}=h_{nn_b,0}$ in \eqref{EQN_CSIhnt}.
Thus, we only need to apply \eqref{Eqn:IRS:phaseNoise} for \eqref{EQN_CSIhnt} to model the varied phase noise, resulting in
\begin{equation}  \label{EQNIrs_gainIRSUEslott}
\mathfrak{h}_{nn_b,t} = h_{nn_b,t} e^{j\phi_{t}}= h_{nn_b,0} e^{j(\phi_{0} + \triangle \Phi_{t})}.
\end{equation}
Substituting \eqref{EQNIrs_gainIRSUEslott} into \eqref{EQNirs_vecIRSUE}, we obtain
\begin{equation}
    \mathbf{h}_{n,t}=\mathbf{h}_{n,0}e^{j\triangle \Phi_{t}},\: \text{and}\:\: \mathbf{H}_{t}=\mathbf{H}_{0}e^{j\triangle \Phi_{t}}.
\end{equation}

\section{Performance Analysis}
\begin{figure*}[!t]
\setcounter{equation}{20}
\begin{align} \label{eqn:longIRSRx} \nonumber
 \gamma_t&=\frac{P_d }{\sigma_n^2}  \left| \left( \sum_{n=1}^N \left[\rho_t \mathfrak{g}_{n,0} e^{j\triangle \Psi_{t}}+ \varepsilon_n \sqrt{1-\rho_t^2}   e^{j(\psi_{0}+\triangle \Psi_{t})} \right] e^{j(\theta_n^\star+\tilde{\theta}_n)} \mathbf{h}_{n,0}^Te^{j\triangle \Phi_{t}} +  \left[ \rho_t\mathbf{h}_{d,0}e^{j\left(\triangle \Phi_{t}+\triangle \Psi_{t}\right)}+ \boldsymbol{\varepsilon}_d \sqrt{1-\rho_t^2}   e^{j\left(\phi_{0}+\psi_{0}+\triangle \Phi_{t}+\triangle \Psi_{t}\right)} \right]^T \right)\mathbf{w}^\star     \right|^2 \\&=\frac{P_d }{\sigma_n^2}  \left| \left( \left[\rho_t \mathbf{g}_{0} e^{j\triangle \Psi_{t}}+ \boldsymbol{\varepsilon}_g \sqrt{1-\rho_t^2} e^{j(\psi_{0}+\triangle \Psi_{t})}   \right]^T \Theta^\star \tilde{\Theta} \mathbf{H}_{0}e^{j\triangle \Phi_{t}}  +  \left[  \rho_t\mathbf{h}_{d,0}e^{j\left(\triangle \Phi_{t}+\triangle \Psi_{t}\right)}+ \boldsymbol{\varepsilon}_d \sqrt{1-\rho_t^2} e^{j\left(\phi_{0}+\psi_{0}+\triangle \Phi_{t}+\triangle \Psi_{t}\right)}  \right]^T  \right )\mathbf{w}^\star \right|^2
\end{align} \rule{\textwidth}{0.1mm}
\begin{align} \label{eqn:longIRSRxSimplified}
 \gamma_t&=\frac{P_d }{\sigma_n^2}  \left| \left( \left[\rho_t \mathbf{v}_{g_0} + \boldsymbol{\varepsilon}_g \sqrt{1-\rho_t^2}    \right]^T \Theta^\star \tilde{\Theta} \mathbf{V}_{H_0}e^{j(\phi_0+\psi_{0}+\triangle \Psi_{t}+\triangle \Phi_{t})}  +  \left[  \rho_t\mathbf{v}_{d_0}e^{j\left(\phi_{0}+\psi_{0}+\triangle \Phi_{t}+\triangle \Psi_{t}\right)}+ \boldsymbol{\varepsilon}_d \sqrt{1-\rho_t^2} e^{j\left(\phi_{0}+\psi_{0}+\triangle \Phi_{t}+\triangle \Psi_{t}\right)}  \right]^T  \right )\mathbf{w}^\star \right|^2\\ \label{eqn:yx2}
 &=\frac{P_d }{\sigma_n^2}  \left|  \left[\rho_t \mathbf{v}_{g_0} + \boldsymbol{\varepsilon}_g \sqrt{1-\rho_t^2}     \right]^T \Theta^\star \tilde{\Theta} \mathbf{V}_{H_0}  +  \left[  \rho_t\mathbf{v}_{d_0}+ \boldsymbol{\varepsilon}_d \sqrt{1-\rho_t^2}   \right]^T   \right|^2
\end{align}  \setcounter{equation}{19}
\rule{\textwidth}{0.1mm}
\end{figure*}

The BS estimates $\mathbf{g}_0$, $\mathbf{H}_0$, and $\mathbf{h}_{d,0}$ via the uplink training at slot $0$, whereas it does not know the actual CSI $\mathbf{g}_t$, $\mathbf{H}_t$, and $\mathbf{h}_{d,t}$ at slot $t=1,\ldots,\mathcal{T}$ \footnote{Despite of focusing on a single-user IRS system,  as most of the previous works \cite{Ref_wu2019intelligent, Ref_you2020channel, Ref_han2019large}, for ease of illustration, the analyses and evaluation are applicable to a multi-user IRS system without any need of major modifications.  With the aid of orthogonal multiple access techniques, such as TDMA or OFDMA/FDMA, each user assigned to an orthogonal time slot or frequency sub-carrier is exactly equivalent to the single user analyzed in the letter, where the modelling of channel aging/phase noise and the alternative optimization can be employed straightforwardly. }.
Hence, it can only optimize the IRS transmission based on the outdated information. By jointly designing the active beamforming $\mathbf{w}$ and passive reflection coefficients $\boldsymbol{\Theta}$, the BS aims to maximize $R_0=\log_2\left(1+\frac{P_d}{\sigma_n^2}\left|(\mathbf{g}_{0}^T \boldsymbol{\Theta} \mathbf{H}_0 +\mathbf{h}_{d,0}^T)\mathbf{w}\right|^2\right)$, resulting in the following optimization problem
\begin{equation}
\begin{aligned} \label{eqnIRS:optimizationMRTvector}
\max_{\boldsymbol{\Theta},\:\mathbf{w}}\quad &  \biggl|\Bigl(\mathbf{g}_0^T \boldsymbol{\Theta} \mathbf{H}_0 +\mathbf{h}_{d,0}^T\Bigr)\mathbf{w}\biggr|^2\\
\textrm{s.t.} \quad & \|\mathbf{w}\|^2\leqslant 1\\
  \quad & \theta_n\in [0,2\pi), \: \forall n=1,2,\ldots,N.
\end{aligned}
\end{equation}
Alternating optimization can be applied to solve this non-convex problem, achieving near-optimal performance with affordable complexity on the order of magnitude $\mathcal{O}(N^2N_b)$. The iterative optimization process is not hard to derive from \cite{Ref_wu2019intelligent}, which is omitted due to the page limit. This process iterates until the convergence is reached with the optimal transmit vector $ \mathbf{w}^\star = \frac{\left(\mathbf{g}_0^T \boldsymbol{\Theta}^{\star} \mathbf{H}_0 +\mathbf{h}_{d,0}^T\right)^H}{\|\mathbf{g}_0^T \boldsymbol{\Theta}^{\star} \mathbf{H}_0 +\mathbf{h}_{d,0}^T\|}$ and phase-shift matrix $\boldsymbol{\Theta}^\star$.

Substituting $\mathbf{w}^\star$ and $\boldsymbol{\Theta}^\star$ into \eqref{eqn_systemModel}, yields the received signal at time slot $t$ and the received signal-to-noise ratio (SNR), which is given by \eqref{eqn:longIRSRx} at the top of this page. Correspondingly, the achievable spectral efficiency at time slot $t$ is computed by $R_t=\log_2\left(1+\gamma_t\right)$, and the average spectral efficiency of a radio frame equals $\bar{R}=\frac{1}{T+1}\sum_{t=1}^T R_t$.

Interestingly, it is observed that the oscillator phase noise does not affect the performance of an IRS-aided communication system.  Let's elaborate this feature mathematically as follows:
\newtheorem{theorem}{Theorem}
\begin{theorem}
\label{Lemma_02}
The instantaneous SNR $\gamma_t$ given in \eqref{eqn:longIRSRx} is independent of $\phi_t$ and $\psi_t$.
\end{theorem}
\begin{IEEEproof}
Substituting $t=0$ into \eqref{EQN_CSIgnt} and \eqref{EQNIRS:VectorIRSUE} yields
\setcounter{equation}{23}
\begin{equation}  \label{eqn_CSIg0}
    \mathbf{g}_0 =\left[\mathfrak{g}_{1,0},\ldots,\mathfrak{g}_{N,0}\right]^T=e^{j\psi_{0}}\cdot\left[g_{1,0},\ldots,g_{N,0}\right]^T=e^{j\psi_{0}} \mathbf{v}_{g_0},
\end{equation}
where $\mathbf{v}_{g_0}=\left[g_{1,0},\ldots,g_{N,0}\right]^T$. Defining a matrix $\mathbf{V}_{H_0}$ that consists of entries $h_{nn_b,0}$ and $\mathbf{v}_{d_0}=[h_{d1,0},\ldots,h_{dN_b,0}]^T$, we get
\begin{equation} \label{eqn_x2}
    \mathbf{h}_{d,0}=e^{j(\phi_{0}+\psi_{0})}\mathbf{v}_{d_0},\:\text{and}\:\:
    \mathbf{H}_0=e^{j\phi_0}\mathbf{V}_{H_0}
\end{equation} by recalling \eqref{EQN_CSIhd0}, \eqref{EQN_CSIhnt}, \eqref{EQNIRS_VectorBSUE}, and \eqref{EQNirs_vecIRSUE}.
Substitute \eqref{eqn_CSIg0} and \eqref{eqn_x2} into \eqref{eqn:longIRSRx} to obtain \eqref{eqn:longIRSRxSimplified}, where the first term gets a phase error of $e^{j\left(\phi_{0}+\psi_{0}+\triangle \Phi_{t}+\triangle \Psi_{t}\right)}$, while the second term suffers from an equivalent noise. Hence, the phase alignment between the direct  and  reflected components remains.
Similarly, the optimal transmit vector can be rewritten as
\begin{equation} \label{EQN_BestBFvector}
    \mathbf{w}^{\star} = \frac{\left(\mathbf{g}_0^T \boldsymbol{\Theta}^{\star} \mathbf{H}_0 +\mathbf{h}_{d,0}^T\right)^H}{\left\|\mathbf{g}_0^T \boldsymbol{\Theta}^{\star} \mathbf{H}_0 +\mathbf{h}_{d,0}^T\right\|}=\frac{e^{j(\phi_{0}+\psi_{0})}\left(\mathbf{v}_{g_0}^T \boldsymbol{\Theta}^{\star} \mathbf{V}_{H_0} +\mathbf{v}_{d_0}^T\right)^H}{\left\|\mathbf{v}_{g_0}^T \boldsymbol{\Theta}^{\star} \mathbf{V}_{H_0} +\mathbf{v}_{d_0}^T\right\|}.
\end{equation}
Then, \eqref{eqn:longIRSRxSimplified} is further simplified to \eqref{eqn:yx2}, where the effect of oscillator phase noise vanishes since $\left| e^{j\left(\phi_{0}+\psi_{0}+\triangle \Phi_{t}+\triangle \Psi_{t}\right)}\right|=1$ and $\left| e^{j\left(\phi_{0}+\psi_{0}\right)}\right|=1$\footnote{Some previous IRS works consider the scenario without a BS-UE link. By setting $h_{dn_b,t}=0$ or $\mathbf{h}_d=\mathbf{0}$, the general setup of this letter is simplified to this special case. Correspondingly, its closed-form expression of the received SNR is obtained by removing the second term of \eqref{eqn:yx2}. It is not hard to know from the derivation process that the oscillator phase noise also does not affect the performance when there is no BS-UE link. }.
\end{IEEEproof}

\begin{figure*}[!tbph]
\centerline{
\subfloat[]{
\includegraphics[width=0.27\textwidth]{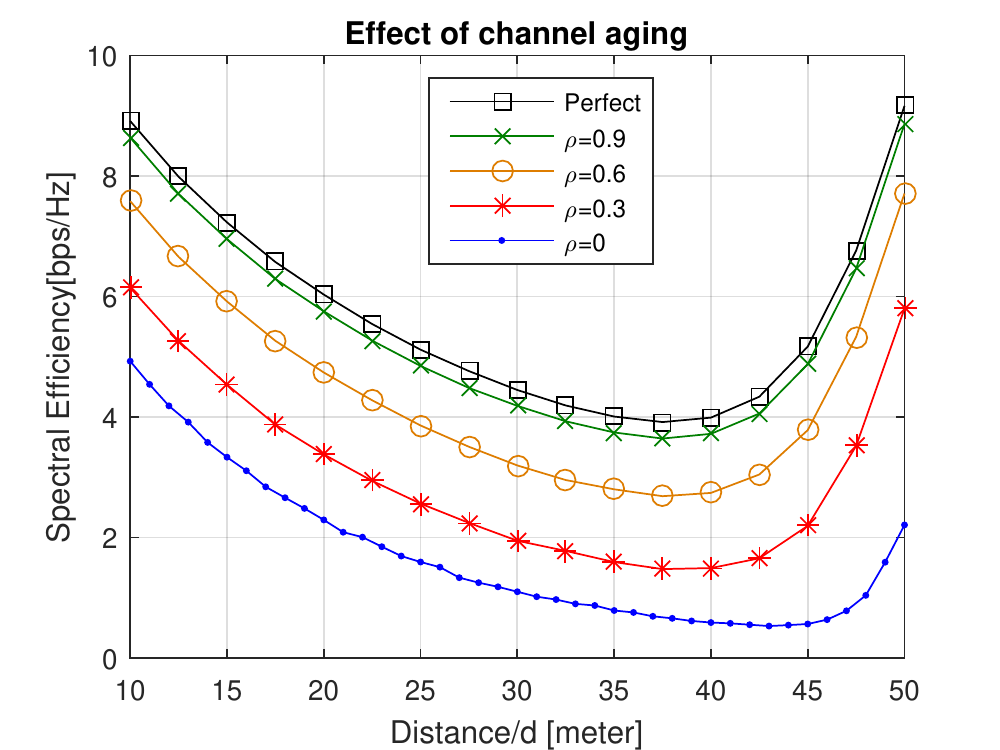}
\label{fig:result1}
}
\hspace{-8mm}
\subfloat[]{
\includegraphics[width=0.27\textwidth]{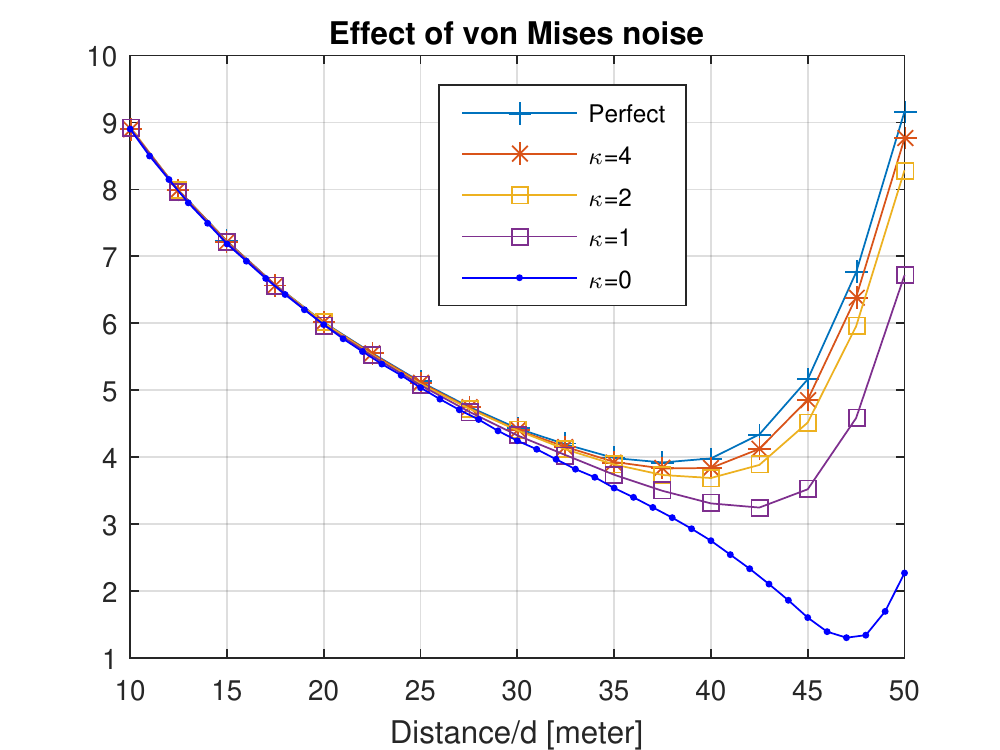}
\label{fig:result2}
}
\hspace{-8mm}
\subfloat[]{
\includegraphics[width=0.27\textwidth]{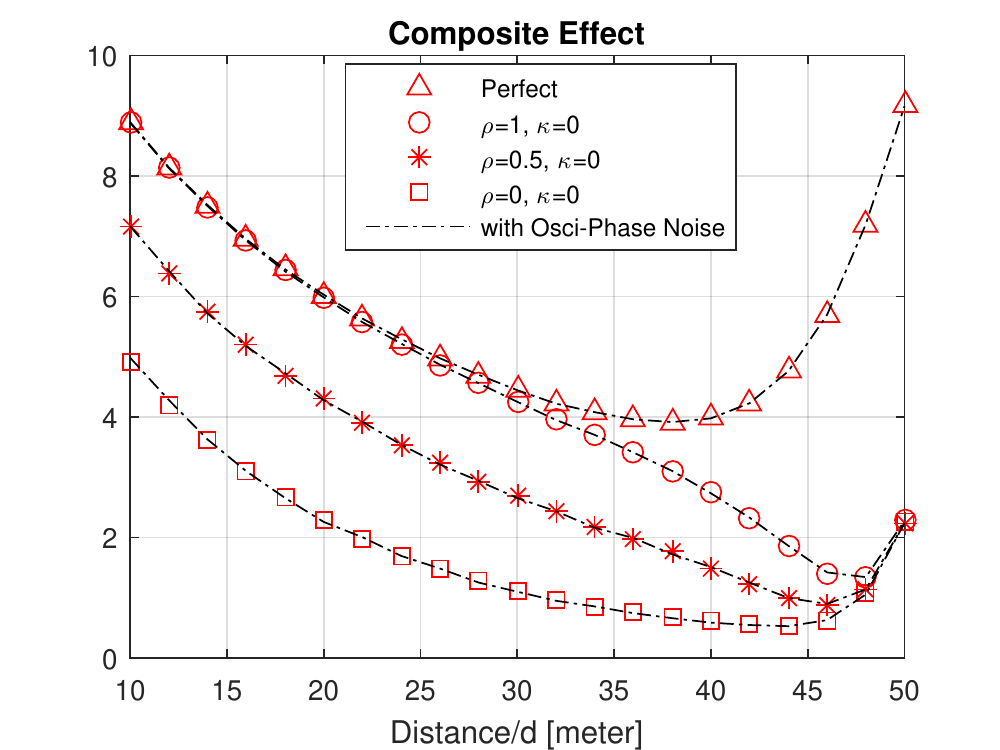}
\label{fig:result3}
}
\hspace{-8mm}
\subfloat[]{
\includegraphics[width=0.27\textwidth]{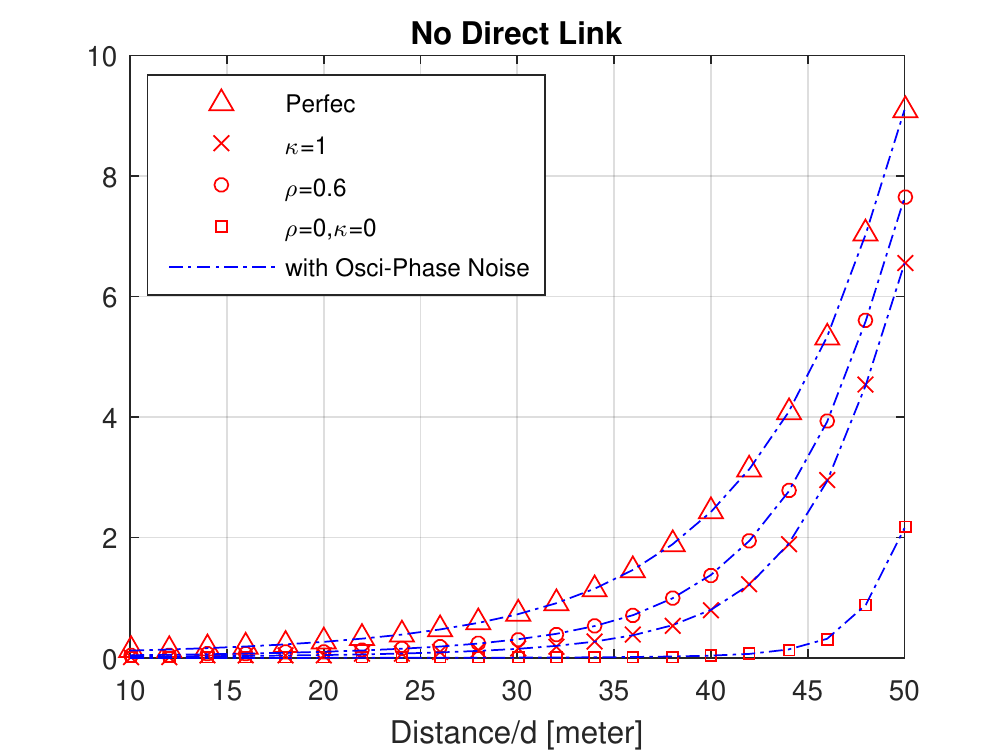}
\label{fig:result4}
}
}
\hspace{0mm}
 \caption{Spectral efficiencies of an IRS-aided system under channel and hardware impairments:  (a) the effect of channel aging in terms of correlation coefficients from $\rho=1$ to $\rho=0$; (b) the effect of IRS phase noise from $\kappa=0$ to $\kappa=+\infty$; (c) the composite effect including oscillator phase noise (Osci-Phase Noise); and (d) the composite effect under the scenario without a direct link.  }
\label{Fig_performance}
\end{figure*}

\section{Numerical results}
This section explains the simulation setup and provides some typical numerical examples to demonstrate the performance impact of channel and hardware impairments on the IRS-aided wireless systems. For the sake of comparison, we use the same three-dimensional coordinate system in \cite{Ref_wu2019intelligent}, as given by its \figurename 2.
The large-scale fading is calculated by $L(d)=L_0/d_0^{-\alpha}$, where $L_0$ is the path loss at the reference distance of \SI{1}{\meter}, $d_0$ stands for the propagation distance, and $\alpha$ means the path-loss exponent. Taking into account the shadowing, an extra loss of \SI{10}{\decibel} is applied for both the BS-UE and IRS-UE links, whereas the BS-IRS link is usually LOS without blockage in-between. Meanwhile, the Rician channel model is used for small-scale fading, i.e.,
\begin{equation}
    \mathbf{h}=\sqrt{\frac{K}{K+1}}\mathbf{h}_{LOS} + \sqrt{\frac{1}{K+1}}\mathbf{h}_{NLOS},
\end{equation}
where  $K$ denotes the Rician factor, $\mathbf{h}_{LOS}$ is the LOS component, and $\mathbf{h}_{NLOS}$ stands for the multipath component.
The BS and IRS are apart from $d_{BI}=51\si{\meter}$ and the UE lies on a horizontal line with the vertical distance of $d_v=2\si{\meter}$. The horizontal distance between the BS and UE is denoted by $d$. Accordingly, the BS-UE and  IRS-UE distances are computed by $d_{BU}=\sqrt{d^2+d_v^2}$ and $d_{IU}=\sqrt{(d_{BI}-d)^2+d_v^2}$, respectively. Other simulation parameters are $N=200$, $N_b=16$, $L_0=\SI{-30}{\decibel}$, $P_d=5\mathrm{dBm}$, $\sigma_n^2=-80\mathrm{dBm}$, the Rician factor and path-loss exponent for the direct, BS-IRS, and IRS-UE links are  $\{K=0$, $\alpha=3\}$, $\{K=+\infty$, $\alpha=2\}$, and $\{K=0$, $\alpha=3\}$, respectively. For ease of exposition, we simply set $T=1$ in the simulations to demonstrate the impact of channel aging and phase noise. Since the ranges of parameters are set to be large enough, i.e., $\rho\in[0,1]$ and $\kappa\in[0,+\infty)$, the results obtained from $T=1$ are sufficiently representative for the purpose of observing the performance impact.

The results of spectral efficiency as a function of $d$ are provided in \figurename \ref{Fig_performance}.
We use the performance of the alternating optimization without channel aging and phase noise, where $\rho=1$, $\kappa\rightarrow \infty$, and $\phi_t=\psi_t=0$, as the benchmark, which is denoted by \textit{Prefect} in the figures. The alternating optimization achieves the optimal performance for the joint active and passive beamforming as revealed by \cite{Ref_wu2019intelligent}. In our simulations, the number of iterations is set to three, which is enough for convergence. In a conventional system without IRS, the cell-edge user suffers from low SNR due to severe propagation loss. With the aid of an IRS, cell-edge performance can be improved because the far user can get extra reflected signals. If the BS-IRS connection is properly established and strong inter-cell interference is avoided, the performance curve is U-shape, as demonstrated by Fig. 3 of \cite{Ref_wu2019intelligent}.

\figurename \ref{fig:result1} shows the spectral-efficiency results as a function of the horizontal distance $d$ between the BS and UE under different extents of channel aging. At $\rho=0.9$, the performance loss is still trivial, around \SI{-0.2}{\bps\per\hertz^{}}. The loss is enlarged to approximately \SI{-1.1}{\bps\per\hertz^{}} and \SI{-2.2}{\bps\per\hertz^{}} when $\rho$ decreases to $0.6$ and $0.3$, respectively. The worse case is $\rho=0$, which means the measured CSI is completely uncorrelated with the actual CSI, imposing a substantial loss of over \SI{-4}{\bps\per\hertz^{}}. \figurename \ref{fig:result2} reveals the impact of IRS phase noise in terms of different values of $\kappa$.  In the case of $\kappa=0$, where each reflecting element suffers from the worst phase noise that equally distributes over $[0,2\pi)$,  the cell-edge performance degrades around \SI{7}{\bps\per\hertz^{}}. If $\kappa>1$, the loss becomes moderate with about \SIrange{-1}{-2}{\bps\per\hertz^{}}. As we expected, it is observed that IRS phase noise mainly affects the cell-edge users since the IRS locates at the cell edge in this simulation. The IRS phase noise does not affect the active BS beamforming, and the impact on the cell-center users is negligible. In contrast, the difference between \figurename \ref{fig:result1} and \figurename \ref{fig:result2} highlights the fact that the channel aging degrades both the performance of active beamforming and passive beamforming.

\figurename \ref{fig:result3} illustrates the composite effects including oscillator phase noise.  In this figure, the markers denote the  spectral efficiencies without oscillator phase noise by setting $\phi_t=0$ and $\psi_t=0$, while the dash-dotted lines stand for the results with oscillator phase noise but other conditions are unchanged. It is observed that the oscillator phase noise does not affect the performance  at different values of $\rho$ and $\kappa$, justifying the correctness of Theorem 1. In the worst case, which has the most aged CSI $\rho=0$ and the worst IRS phase noise $\kappa=0$, the system has the lower performance bound. It is comparable to the curve of $\rho=0$ in \figurename \ref{fig:result1}, implying that the IRS-aided systems suffer mostly from the channel aging, and the IRS phase noise is negligible under aged channels. Last but not least, \figurename \ref{fig:result4} illustrates the impact of these impairments under the scenario without a direct link. With only the reflected signals, the cell center becomes a dark spot, where the users suffer from poor performance. In contrast, the cell-edge users get good quality of service. The impact of channel aging with $\rho=0.6$ and reflector phase noise with $\kappa=1$ are demonstrated in the figure, in comparison with the lower bound ($\rho=0$ and $\kappa=0$). In this case, the oscillator phase noise still does not affect the system performance, justifying Theorem 1.

\section{Conclusions}
This letter theoretically analyzed and numerically evaluated the impact of channel aging and phase noise on IRS-aided wireless systems. It revealed that oscillator phase noise \textit{does not affect} the performance because the phase alignment between the direct and reflected signals is still kept. Therefore, we do not need to consider  oscillator imperfection when designing an IRS system. In contrast, we should deal with the effect of high Doppler shifts and IRS hardware impairment, which substantially deteriorate the performance. Despite using a single-cell, single-user setup, the mathematical models and analyses developed in this letter can be extended to multi-user, multi-cell scenarios without major revisions.

\bibliographystyle{IEEEtran}
\bibliography{IEEEabrv,Ref_COML}

\end{document}